\documentclass[aps,twocolumn,showpacs,superscriptaddress,preprintnumbers,amsmath,amssymb]{revtex4}
\usepackage{graphicx}
\usepackage{dcolumn}
\usepackage{bm}
\usepackage{color}

\begin{document}

\title{Impact of two-electron dynamics and correlations \\ on high-order harmonic generation in He}

\author{\firstname{Anton~N.} \surname{Artemyev}}
\affiliation{Institut f\"{u}r Physik und CINSaT, Universit\"{a}t Kassel, Heinrich-Plett-Str. 40, 34132 Kassel, Germany}

\author{\firstname{Lorenz~S.} \surname{Cederbaum}}
\affiliation{Theoretische Chemie, Physikalisch-Chemisches Institut, Universit\"{a}t Heidelberg, Im Neuenheimer Feld 229, 69120 Heidelberg, Germany}
\date{\today}

\author{\firstname{Philipp~V.} \surname{Demekhin}}\email{demekhin@physik.uni-kassel.de}
\affiliation{Institut f\"{u}r Physik und CINSaT, Universit\"{a}t Kassel, Heinrich-Plett-Str. 40, 34132 Kassel, Germany}
\affiliation{Research Institute of Physics, Southern Federal University, Stachki\,av.\,194, 344090 Rostov-on-Don, Russia}

\date{\today}

\begin{abstract}
The interaction of a helium atom with intense short 800\,nm laser pulse is studied theoretically beyond the single-active-electron approximation. For this purpose, the time-dependent Schr\"odinger equation for the two-electron wave packet driven by a linearly-polarized infrared pulse is solved by the time-dependent restricted-active-space configuration-interaction method (TD-RASCI) in the dipole velocity gauge. By systematically extending the  space of active configurations, we investigate the role of the collective two-electron dynamics in the strong field ionization and high-order harmonic generation (HHG) processes. Our numerical results demonstrate that  allowing both electrons in He to be dynamically active results in a considerable extension of the computed HHG spectrum.
\end{abstract}


\maketitle

\section{Introduction}
\label{sec:intro}

An electron released from an atom or molecule and further on steered in the field of the parent ion by intense laser fields gives rise to many fundamental phenomena \cite{Corkum93}. For instance, the field driven electron-ion recombination results in the emission of coherent radiation with frequencies which are odd integer multiples of the carrier frequency of exciting pulse, a phenomenon known as the high-order harmonic generation (HHG) process. Over the last two decades, HHG has been intensively studied experimentally and theoretically (see, e.g., review articles \cite{Rev1,Rev2,Rev3} and references therein), owing to its powerful application to the generation of coherent XUV laser pulses down to the attosecond regime  \cite{Atto1,Atto2,Atto3,Atto4}.

In principle, high harmonics can be generated by transitions among excited bound electronic states as well as by the involvement of unbound continuum electronic states. Detailed calculations on an array of quantum dots with only bound discrete electronic states have demonstrated an efficient HHG \cite{nimrod}. However, as the current experiments on atoms and molecules are performed in the strong-field ionization regime, below we concentrate on this regime. Here, the essential mechanism behind the HHG process is explained within the simplified three-step model \cite{Corkum93,TSM1,TSM2}, in which an electron: (i) escapes from the potential formed by the superposition of the ionic core and linearly-polarized laser field potentials; (ii) is accelerated back to the parent ion during the next half-cycle of the driving pulse; and (iii) recombines with the ion emitting thereby high-energy photons. The validity of this illustrative model has been confirmed by numerous theoretical studies of HHG process  in atoms and molecules performed within the single-active-electron (SAE) approximation (for recent results see, e.g., Refs.~\cite{SAE1,SAE2,SAE3,SAE4,SAE5,SAE6,SAE7,SAE8,SAE9} and references therein).

The role of inactive electrons, which are kept frozen within the SAE approximation, is not negligible  for the strong-field processes. However, exact numerical solution of the time-dependent Schr\"odinger equation (TDSE) for many-electron systems  in laser fields is very formidable task. At present, it has been realized only for He atom \cite{Hesf1,Hesf2,Hesf3,Hesf4,Hesf5,Hesf6,Hesf7} to calculate single- and double-electron  multi-photon ionization rates. The method of full dimensional numerical integration of TDSE for two electrons, developed in these works, utilizes a basis set of coupled spherical harmonics \cite{Hesf1}, and it is particularly efficient to study strong-field problems which can be described by implying very small radial grids of about 100~Bohr.

There are several theoretical studies of HHG process performed beyond the SAE approximation in atoms \cite{aCI1,aCI2,aCI3,aCI4} and molecules \cite{mCI1,mCI2,mCI3,mCI4,mCI5}. Ref.~\cite{aCI1}, for instance, proposes generalization of the three-step model  for many-electron systems by introducing perturbative corrections due to exchange and electron correlations. A more consistent tracking of electron correlations and dynamics is provided by the multi-configuration time-dependent Hartree-Fock (MCTDHF) method \cite{MCTDHF1,MCTDHF2,MCTDHF3,MCTDHF4,MCTDHF5,MCTDHF6}. Its straightforward implementation to the solution of the strong-field problems is, however, a challenging computational task \cite{MCTDHFapp1,MCTDHFapp2}. In order to be able to study HHG by MCTDHF method in realistic systems, one can either lower dimensionality of the problem \cite{mCI4} (i.e., consider only one or two dimensions for each electron), or/and neglect the exchange interaction \cite{aCI2} (i.e., utilize the MCTDH method \cite{MCTDH}).

Alternatively, relaxing the full configuration interaction character of MCTDHF by cleverly selecting and incorporating only the important electron correlation and dynamical effects into the ansatz for the wavepacket, one can fully catch the essential physics of the process at hand, and at the same time reduce considerably the numerical effort and thus make the computation tractable. This can be realized by limiting the space of active electron configurations utilizing, e.g., the time-dependent restricted-active-space configuration-interaction method (TD-RASCI, \cite{TDRASCI1,TDRASCI2}), the time-dependent generalized-active-space configuration-interaction (TD-GASCI, \cite{mCI5,TDGASCI}), or the time-dependent restricted-active-space self-consistent-field theory (TD-RASSCF, \cite{aCI3}). The latter approach has already been applied to calculate HHG spectra of the 1D beryllium atom \cite{aCI3}.

Recently \cite{TDRASCI2}, we have applied the TD-RASCI method developed in Ref.~\cite{TDRASCI1} to investigate the photoionization of He by intense high-frequency laser pulses. In the present work, we utilize this method to investigate the influence of the correlative electron dynamics on the HHG spectra of this simplest system with two electrons, treated each in three dimensions (i.e., we treat here a six dimensional problem). The paper is organized as follows.  Sec.~\ref{sec:theory} outlines our theoretical approach and justifies the present choice of the active space of electron configurations. In the numerical calculations, we systematically increased the space of active configurations and proceed as far as we could. These numerical results are presented and  discussed in Sec.~\ref{sec:results}. We conclude with a brief summary.

\section{Theory}
\label{sec:theory}

The present theoretical approach is fully described in our previous work \cite{TDRASCI2}. More details on its numerical implementation can be found in  Refs.~\cite{KEoper,FEDVR,FEDVRTR,DIRPROP,PECD}. Therefore, only essential relevant points  are discussed below.

We describe the light-matter interaction in the velocity gauge, which is the most suitable gauge for strong field problems \cite{velocity}, since it ensures rapid convergence of the numerical solution over angular momentum of released photoelectrons \cite{SAE6}. In the electric dipole approximation, the total Hamiltonian governing dynamics of two electrons of He exposed to intense coherent linearly-polarized laser pulse  reads (atomic units are used throughout)
\begin{multline}
\hat{H}(t)= -\frac{1}{2}\vec{\nabla}^2_1 -\frac{1}{2}\vec{\nabla}^2_2 -\frac{2}{r_1}-\frac{2}{r_2} +\frac{1}{\vert \vec{r}_1-\vec{r}_2\vert} \\
-i\left( \nabla_{z_1}+\nabla_{z_2}\right) \mathcal{ A}_0 \,g(t) \sin(\omega t).
\label{eq:Hamlt}
\end{multline}
Here, $g(t)$ is the time-envelope of the pulse, $\omega$ is its carrier frequency,  $\mathcal{ A}_0$ is the peak amplitude of the vector potential (the vector potential and the electric field vector are related via $\mathbf{E}=-\partial_t\mathbf{A}$), and the  peak intensity of the pulse is $I_0 =\frac{\omega^2}{8\pi\alpha}\mathcal{A}_0^2$ ($\alpha\simeq1/137.036$ is the fine structure constant, and 1~a.u. of intensity is equal to 6.43641$\times 10^{15}$ W/cm$^2$).

The present calculations were performed for laser pulses with carrier frequency of $\omega=0.0569$~a.u. (corresponding to a wavelength of $\lambda=800$~nm) and peak intensity of $5\times10^{14}$~W/cm$^2$. For this photon energy, ionization of He requires the absorption of at least 16 photons (the ionization potential is 24.587~eV \cite{NIST}). The corresponding Keldysh \cite{KELD} parameter $\gamma = 0.64$ indicates that ionization takes place in the intermediate regime between the strong-field tunnel-ionization ($\gamma \ll 1$) and multiphoton ionization ($\gamma \gg 1$) extremes. At the chosen field strength, the rates for double ionization of He are very small compared to its single ionization rates \cite{Hesf3,Hesf6,PBUKS}. We may, therefore, neglect in the present study of the HHG process the double ionization and permit only one of
the electrons of He to be ionized by the pulse. Nevertheless, we allow the bound electron to interact with the laser field as well as with the photoelectron, i.e., it is fully active but kept bound.

To accomplish the above description, we utilized the following symmetrized ansatz for the spatial part of the total two-electron wave function $\Psi(\vec{r}_1,\vec{r}_2,t)$  in the singlet spin state:
\begin{multline}
\Psi(\vec{r}_1,\vec{r}_2,t)=\sum_\alpha a_\alpha(t)\phi_\alpha(\vec{r}_1)\phi_\alpha(\vec{r}_2)\\
+\sum_{\alpha > \alpha^\prime} b_{\alpha \alpha^\prime}(t)\frac{1}{\sqrt {2}}\left[\phi_\alpha(\vec{r}_1)\phi_{\alpha^\prime}(\vec{r}_2)+ \phi_{\alpha^\prime}(\vec{r}_1)\phi_\alpha(\vec{r}_2) \right]\\
+\sum_{\alpha\beta} \frac{1}{\sqrt {2}}\left[\phi_\alpha(\vec{r}_1)\psi_{\beta}(\vec{r}_2,t)+ \psi_{\beta}(\vec{r}_1,t)\phi_\alpha(\vec{r}_2) \right].
\label{eq:tewf}
\end{multline}
As justified above, the wave function (\ref{eq:tewf}) is constructed by using two different mutually-orthogonal one-electron spatial basis sets $\left\{ \phi_{\alpha}(\vec{r}\,)\right\}$ and $\left\{ \psi_{\beta} (\vec{r},t) \right\}$. The former describes dynamics of the electron which remains bound to the nucleus, and it includes selected discrete orbitals $\left\{\phi_{\alpha}(\vec{r}\,)\equiv \phi_{n\ell m} (\vec{r}\,)\right\}$. The latter is formed by the time-dependent wave packets of a photoelectron $\left\{ \psi_{\beta} (\vec{r},t)\equiv  \psi^\alpha_{\ell m} (\vec{r},t) \right\}$,  which  are built to be orthogonal to all incorporated discrete orbitals, i.e., $\langle \phi_{\alpha} \vert \psi_{\beta }(t) \rangle=0$.

In order to describe these one-electron basis sets, we applied the finite-element discrete-variable representation (FEDVR) scheme and introduced the three-dimensional basis element $\xi_{\lambda} (\vec{r}\,)$ as:
\begin{equation}
\label{eq:basis3d}
\xi_{\lambda} (\vec{r}\,) \equiv \xi_{ik,\ell m} (\vec{r}\,)=\frac{\chi_{ik}(r)}{r}\,Y_{\ell m}(\theta,\varphi).
\end{equation}
Here, the radial coordinate is represented by the  basis set of the normalized Lagrange polynomials $\chi_{ik}(r)$  \cite{KEoper,FEDVR,FEDVRTR,DIRPROP,PECD} constructed over a Gauss-Lobatto grid $\left\{ r_{ik}\right\}$ (index $i$ runs over the finite intervals $\left[r_i,r_{i+1} \right]$ and index $k$ counts the basis functions in each interval). Using this FEDVR, the normalized stationary orbitals and  the time-dependent wave packets  can be  expanded  as follows (note that $\lambda\equiv \left\{ik,\ell m\right\}$ is four-dimensional index):
\begin{subequations}
\label{eq:basis}
\begin{equation}
\label{eq:phibas}
\phi_{\alpha}(\vec{r}\,) =\sum_\lambda d^{\,\alpha}_\lambda\, \xi_{\lambda}(\vec{r}\,),
\end{equation}
\begin{equation}
\label{eq:psibas}
\psi_{\beta}(\vec{r},t) =\sum_\lambda c^{\,\beta}_\lambda(t) \, \xi_{\lambda}(\vec{r}\,).
\end{equation}
\end{subequations}

In the used basis set of the three-dimensional elements Eq.~(\ref{eq:basis3d}), all matrix elements of the Hamiltonian (\ref{eq:Hamlt}) can be evaluated analytically. The corresponding explicit expressions can be found in our previous work \cite{TDRASCI2}, apart from the  light-matter interaction term which was treated there in the dipole length gauge. In the  velocity gauge used here, the dipole transition matrix element reads
\begin{widetext}
\begin{multline}
\label{eq:dipvel}
\left\langle \xi_\lambda \left|\nabla_{z}\right| \xi_{\lambda^\prime}\right\rangle=
\left\langle \frac{\chi_{ik}}{r}\,Y_{\ell m}\left| \cos\theta \frac{\partial}{\partial r} -
\frac{\sin\theta}{r}\frac{\partial}{\partial \theta} \right|\frac{\chi_{i^\prime k^\prime}}{r}\,Y_{\ell^\prime m^\prime}\right\rangle= \left[
\sqrt{\frac{(2\ell+3)(\ell-m+1)(\ell+m+1)}{2\ell+1}}\delta_{\ell^\prime,\ell+1} \frac{\delta_{i,i^\prime}\delta_{k,k^\prime}}{r_{ik}} \right. \\
\left.+\left(\sqrt {\frac{(\ell-m+1)(\ell+m+1)}{(2\ell+3)(2\ell+1)}} \delta_{\ell^\prime,\ell+1}  +\sqrt
{\frac{(\ell-m)(\ell+m)}{(2\ell+1)(2\ell-1)}}\delta_{\ell^\prime,\ell-1}   \right) \left( \int\limits_0^\infty\chi_{ik}\frac{d}{dr}\chi_{i^\prime k^\prime}dr - (\ell^\prime+1)\frac{\delta_{i,i^\prime}\delta_{k,k^\prime}}{r_{ik}} \right)\right]\delta_{m,m^\prime},
\end{multline}
\end{widetext}
where the first derivative  $d\chi_{ik}/dr$ can be evaluated analytically via Eqs.~(10) from Ref.~\cite{TDRASCI2}.

Evolution of the total wave function (\ref{eq:tewf}) in time is given by the vector of the time-dependent expansion coefficients $\vec{ A}(t)=\left\{a_\alpha(t); b_{\alpha \alpha^\prime}(t); c^{\,\beta}_\lambda(t) \right\}$,  which was propagated according to the Hamiltonian (\ref{eq:Hamlt}):
\begin{equation}
\vec{A}(t+\Delta t)=\exp\left\{-iP{\hat{H}(t)}P\Delta t\right\} \vec{A}(t).
\label{eq:prop}
\end{equation}
The one-particle projector  $P=1-\sum_\alpha \vert \phi_\alpha\rangle \langle \phi_\alpha\vert$ in Eq.~(\ref{eq:prop}) acts on the $\left\{c^{\,\beta}_\lambda(t) \right\}$ subspace and ensures the orthogonality condition $\langle \phi_{\alpha} \vert \psi_{\beta }(t) \rangle=0$. The propagation was carried out by the short-iterative Lanczos method \cite{ALG1}. The initial ground state $\vec{A}(t=0)$  was obtained by the propagation in imaginary time (by relaxation) in the absence of the field. The total three-dimensional  photoemission probability was computed as the Fourier transformation of the final electron wave packets at the end of laser pulse:
\begin{equation}
\label{eq:fourier}
W(\vec{k}\,) =\frac{1}{(2\pi)^{3/2}}\sum_\beta \left\vert  \int \psi_{\beta}(\vec{r}) \,e^{-i\vec{k}\cdot\vec{r}} d^3\vec{r}\,  \right\vert ^2.
\end{equation}
Finally, the HHG spectrum $I(\omega)$ was computed as the  squared modulus of the Fourier transformed acceleration of the total electric dipole moment:
\begin{equation}
\label{eq:hhg}
I(\omega) =\frac{1}{(2\pi)^{1/2}} \left\vert  \int  \frac{ d^2 D(t)}{dt^2} \,e^{-i\omega t} dt\,  \right\vert ^2,
\end{equation}
with  $D(t)$ given by
\begin{equation}
\label{eq:dipole}
D(t)=\left\langle \Psi(\vec{r}_1,\vec{r}_2,t) \left \vert \vec{r}_1+\vec{r}_2  \right \vert \Psi(\vec{r}_1,\vec{r}_2,t) \right\rangle .
\end{equation}

The present study was conducted for two different pulse envelopes $g(t)$. The main set of calculations was performed  for a trapezoidal pulse with a linearly-growing front-edge, a constant plateau with unit height, and a linearly-falling back-edge, each supporting five optical cycles. The propagation was thus performed in the time interval of $\left[0,T_f\right]$ with  $T_f=15\frac{2\pi}{\omega}\simeq 40$~fs. In addition, we performed a few key calculations for a sine-squared pulse of the same length  $g(t)=\sin^2(\pi\frac{t}{T_f})$. The size of the radial box was chosen to be $R_{max}=3500$~a.u. The interval $\left[0,R_{max}\right]$  was represented by 1750 equidistant finite elements of the 2~a.u. size, each covered by 10 Gauss-Lobatto points. The photoelectron wave packets were described by the partial harmonics with $\ell \le 50$. Finally, in order to avoid reflection of the wave packets from the boundary at $R_{max}$, they were  multiplied at each time-step by the following mask-function \cite{SAE2,mask}
\begin{equation}
\label{eq:mask}
g(r)= \left\{\begin{array}{ll}1,& r<R_0 \\ \left(\cos \left[ \frac{\pi}{2}  \frac{r-R_0}{R_{max}-R_0} \right]\right)^\frac{1}{8},& R_0<r<R_{max}, \end{array} \right.
\end{equation}
The mask-function (\ref{eq:mask}) was set-in at $R_0=3400$~a.u. Simultaneously, we paid attention that the total electron density in the interval of $\left[0,R_{0}\right]$ does not deviate from its initial value of 2 by more than $10^{-8}$.

\section{Results and discussion}
\label{sec:results}

The present calculations were performed on different levels of approximation which are systematically improved by extending in the total wave function (\ref{eq:tewf}) the set of electronic configurations describing the dynamics of the electron which remains bound. The photoelectron is described fully. For this purpose, in each improvement step we expanded the basis set $\left\{\phi_{\alpha}(\vec{r}\,)\right\}$ describing the bound electron by one additional hydrogen-like $ n\ell_+ $ function of the He$^+$ ion with large quantum numbers  $n$ and $\ell$. The smallest basis set used includes only one $1s_+$ orbital of He$^+$. Hence, in this simplest level of approximation the bound electron is frozen and exhibits no dynamics.  This approximation is an analogue of the SAE approximation. One should stress, however, that it differs from the usual one-electron SAE approximation in hydrogen, since the Hamiltonian (\ref{eq:Hamlt}) includes direct and exchange Coulomb interactions between bound and continuum electrons. The largest basis set of active  orbitals used in the calculations consists of all the discrete one-electron functions $\left\{ n\ell_+ \right\}$ with $n\le4$ and $\ell \le 3$.

\begin{figure}
\includegraphics[scale=.9]{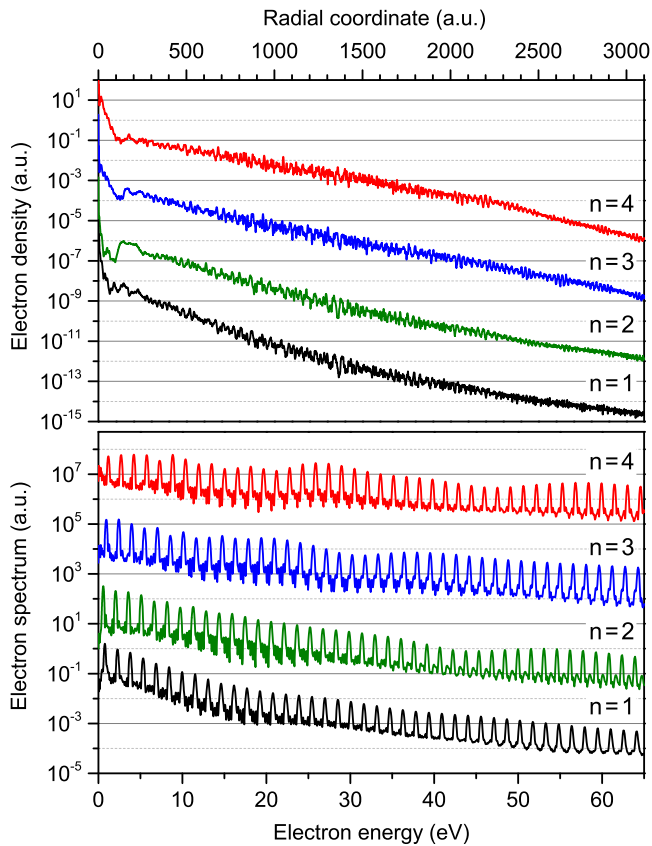}
\caption{(Color online) The final photoelectron radial densities (\textit{upper panel}) and the final photoionization spectra (\textit{lower panel}) after the  linearly-polarized 800\,nm trapezoidal pulse has expired. The time-shape, duration and intensity of the pulse are indicated in the text.  Note the logarithmic scales on the vertical axes. The calculations are performed in a systematic series of improving approximations by sequentially extending the basis set of discrete orbitals $\left\{ n\ell_+ \right\}$, used to describe dynamics of the bound electron. In each step, this basis was extended by the layer of all $n\ell_+$ orbitals with larger principal quantum number $n$, as indicated {at the right-hand side of each spectrum}. The results labeled by {$n=1$} correspond to the SAE approximation. {To enable for a better comparison, the different spectra in each panel are vertically shifted upwards by multiplying successively with $10^3$ starting with the spectrum of the SAE ($n=1$ layer) approximation.}}\label{fig_WpSp}
\end{figure}

We start the present discussion with  analysis of results obtained for the trapezoidal pulse. The upper panel of Fig.~\ref{fig_WpSp} depicts the final radial photoelectron wave packet densities at the end of the laser pulse, computed in the different approximations explained above. The electron density obtained in the SAE approximation (labeled as {$n=1$}) falls exponentially as a function of distance to the nucleus (note the logarithmic scale on the vertical axis). This density is modulated by weak sharp features which represent bunches of fast electrons released by the strong-field multiphoton above-threshold ionization. One can see from this figure that enabling the bound electron in He to be dynamically active causes considerable changes in the computed radial density. Indeed, systematically extending the  $\left\{ n\ell_+ \right\}$ basis set by the layers of orbitals with $n$=2, 3, and 4 results in a significant  lowering of the computed electron density around the nucleus and to its slight enhancement in the outer region. This already indicates the loss of low-kinetic-energy  and gain of high-kinetic-energy electrons in the spectrum.

\begin{figure}
\includegraphics[scale=0.9]{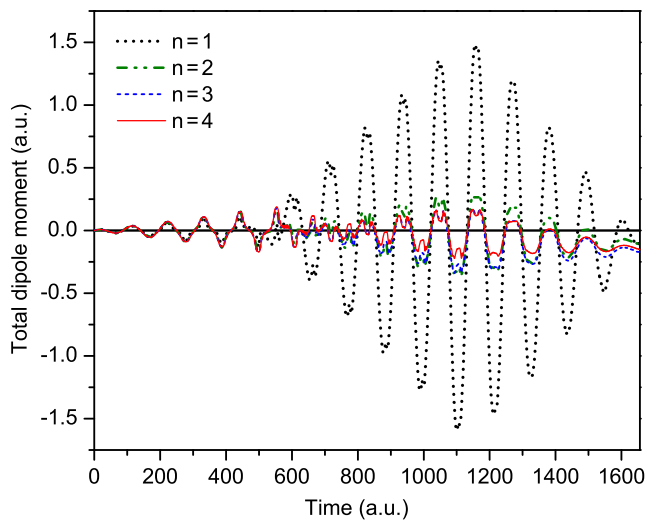}
\caption{(Color online) The total time-dependent electric dipole moment Eq.~(\ref{eq:dipole}), induced by the linearly-polarized 800\,nm trapezoidal pulse described in the text. The calculations are performed in a systematic series of improving approximations as indicated in the caption of Fig.~\ref{fig_WpSp}. }\label{fig_DipMnt}
\end{figure}

The final photoelectron spectra obtained as a Fourier transformation from the final wave packets via Eq.~(\ref{eq:fourier}) are compared in the lower panel of Fig.~\ref{fig_WpSp}. Because of the trapezoidal pulse envelope, the computed spectra consist of a comb of sharp peaks. These  peaks with  exponentially falling intensity are separated by the photon energy $\omega$ (note also the logarithmic scale on the vertical axis). Each peak represents photoelectrons with  kinetic energy of  $\varepsilon_{nj}=E_{n}-E_0-j\omega$ released by the above-threshold multiphoton ionization of He (here $j$ is the number of absorbed photons, $E_0$ is the ground state energy of He, and $E_{n}$ stands for the energy of the $n\ell_+$ ionic state of He$^+$). {By comparing slopes of the electron spectra obtained in different approximation (see lower part of Fig.~\ref{fig_WpSp}),} one can now directly observe the already announced diminution of the low-energy and enhancement of the high-energy parts of the computed spectra, caused by the dynamics of the bound electron.

Figure~\ref{fig_DipMnt} compares the total time-dependent electric dipole moments (\ref{eq:dipole}) computed in different approximations. As is evident from this figure, the impact of the dynamics of the bound electron on this quantity is dramatic: The maximal value of $D(t)$, computed for the largest basis set used here (labeled as {$n=4$}), drops by almost a factor of 7 as compared to the SAE approximation  (labeled as {$n=1$}).

\begin{figure}
\includegraphics[scale=1.4]{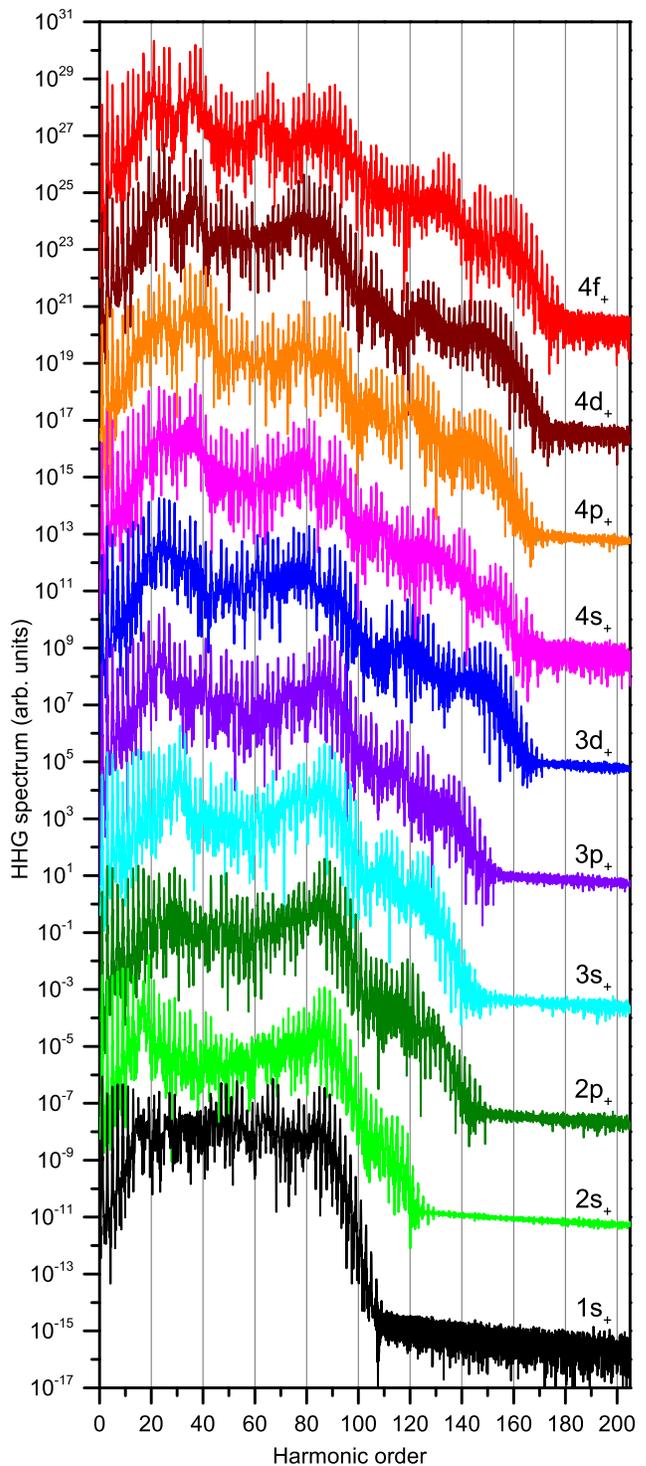}
\caption{(Color online) The HHG spectra of He computed for the trapezoidal laser pulse. The calculations are performed in a systematic series of improving approximations by sequentially extending the basis set of discrete orbitals $\left\{ n\ell_+ \right\}$, used to describe dynamics of the bound electron. In each step, this basis was extended by one orbital, {which is indicated at the right-hand side of each spectrum.} The results labeled by $1s_+$ correspond to the SAE approximation. {To enable for a better comparison, the spectra for each $n\ell$-state are vertically shifted upwards by multiplying successively with $10^4$ starting with the $1s_+$ spectrum.}}\label{fig_HHGtrp}
\end{figure}

We now turn to Fig.~\ref{fig_HHGtrp} which collects all HHG spectra computed for the trapezoidal pulse {  by sequentially extending the basis set of discrete orbitals $\left\{ n\ell_+ \right\}$ by one additional function.} For a better eye view, the spectra  { obtained in each step} are shifted vertically: The {lowermost} spectrum corresponds to the SAE approximation, whereas the  { uppermost one corresponds to the largest basis set used here.}  The HHG spectrum computed in the SAE approximation (the {lowermost} spectrum in Fig.~\ref{fig_HHGtrp} labeled as $1s_+$) exhibits a set of sharp harmonics $k\omega$ with  odd numbers $k \le 100$, which, as expected \cite{Rev1,Rev2,Rev3}, build a typical plateau with a cutoff.

\begin{figure}
\includegraphics[scale=0.9]{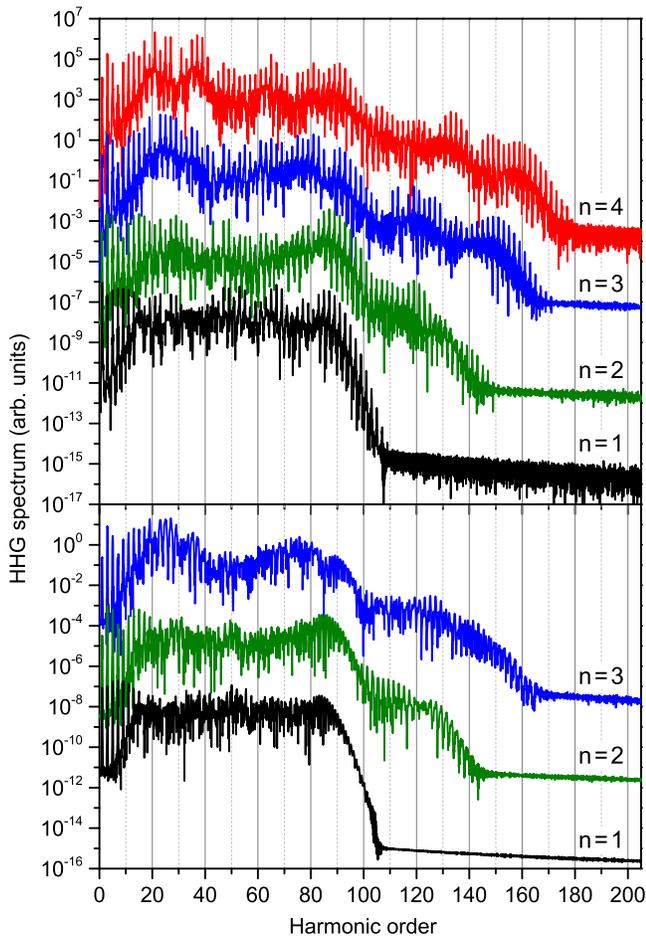}
\caption{(Color online) Summary of the HHG spectra of He, computed for the trapezoidal (\textit{upper panel}) and sine-squared (\textit{lower panel}) laser pulses in different approximations (see caption of Fig.~\ref{fig_WpSp} and text for details). {To enable for a better comparison, the spectra for each $n$-layer are vertically shifted upwards by multiplying successively with $10^4$ starting with the spectrum of the SAE ($n=1$ layer) approximation.}}\label{fig_HHGsin}
\end{figure}

This `classical' picture changes if the dynamics of the bound electron in He is allowed. As one can see from the second from the  {bottom} spectrum, already adding the $2s_+$ state to the basis set of active orbitals for the bound electron extends the number of generated harmonics. Allowing the bound electron to  occupy the $2s_+$ and $2p_+$ orbitals, further extends the number of generated harmonics. One can speak of the formation of a second relatively weak plateau in the HHG spectrum, which starts at the cutoff of the main plateau and exhibits its own cutoff at much higher $k\omega$. Including sequentially the additional $3s_+$, $3p_+$, $3d_+$, and further on the $4s_+$, $4p_+$, $4d_+$, and $4f_+$ orbitals in the  $\left\{ n\ell_+ \right\}$ basis set results in a systematical broadening of the second plateau shifting its cutoff to harmonics  $k\omega$ of higher and higher order $k$ (see Fig.~\ref{fig_HHGtrp}).

The HHG spectra computed by sequentially adding a layer of $\left\{n\ell_+ \right\}$ states with fixed principal quantum number $n=1$, 2, 3, and 4 are summarized again in the upper panel of  Fig.~\ref{fig_HHGsin}.  One can see that the extension of the first plateau of the HHG spectrum does not change as the basis set is enlarged. The extension of the second plateau, which is due to the presence of a second active electron, appears once the basis set contains quantum numbers $n$ larger than 1, and grows fast with additional basis functions {$n=2$ and $n=3$}. The extension of this plateau computed for the largest basis set with {$n = 4$} does, however, not differ much from that obtained with {$n = 3$} indicating a noticeable trend in the convergence of the present computational results with respect to the basis set of discrete orbitals. We stress that the calculations performed here for the largest basis with {$n = 4$} were already at the limit of our computational capabilities.

As a final point of our study, we ensure that the effect observed here is independent of the time-envelope of the laser pulse employed. For this purpose, we performed an analogous set of calculations using a sine-squared pulse of the same length $T_f$ and intensity $I_0$ (for details see the last paragraph of the preceding section). The results of these calculations are collected in the lower panel of Fig.~\ref{fig_HHGsin}. The HHG spectrum computed in the SAE approximation (labeled as {$n=1$}) exhibits a main plateau and cutoff which are very similar to those obtained for the trapezoidal pulse in the same approximation (compare with the  {lowermost spectrum in the upper panel of this figure}). From the lower panel of Fig.~\ref{fig_HHGsin} one can also see that allowing the bound electron in He to occupy the next layer of orbitals with { $n=2$}  results in the formation of a second plateau in the computed HHG spectrum. As demonstrated by the calculations which also include the next layer of orbitals with {$n=3$}, the second cutoff further moves toward higher photon energies.

\section{Conclusion}
\label{sec:concl}
Generation of high-order harmonics in the He atom exposed to intense linearly-polarized 800\,nm laser pulse is studied beyond the single-active-electron approximation by the  time-dependent restricted-active-space configuration-interaction method. During the propagation of the two-electron wave packets in strong laser fields, we allowed only one of the electrons to be ionized and kept the other electron always bound to the nucleus, neglecting thereby the double ionization process which is very weak for the pulse applied. For this purpose, the present active space was restricted to configurations which permit only one of the electrons to populate continuum states. This photoelectron was described in the time-dependent wave packets with angular momenta $\ell \le 50$. The dynamics of the bound electron was described by a set of selected discrete orbitals $\left\{ n\ell_+ \right\}$ of the He$^+$ ion. In the numerical calculations, this discrete one-electron basis was systematically increased by including states with larger quantum numbers  $n$ and $\ell$ up to $n\le4$ and $\ell \le 3$.

The pulse-driven collective correlated dynamics of two electrons in He result in a considerable increase of the number of generated harmonics in the computed HHG spectrum. In particular, compared to the SAE approximation, we observe the formation of a second plateau with weaker harmonics of higher order, which starts at the cutoff of the main plateau and ends with its own cutoff. Increasing sequentially the basis set of active orbitals describing the bound electron results in a systematic extension of the second cutoff to the high-energy side. For the presently used pulse parameters, the computed main plateau extends in the spectrum of harmonics  $k\omega$ up to $k$ of about 100. For the largest basis set of discrete orbitals used here, the second plateau, which is about three orders of magnitude weaker than the main one, consists of additional harmonics with $100<k<180$.

Taking into account that our calculations are probably still not fully converged over the basis set of discrete orbitals
$\left\{ n\ell_+ \right\}$, it is rather difficult to exactly predict the final fate of the second plateau found at higher order harmonics. Nevertheless, the main theoretical conclusion of the present work -- that going beyond the SAE approximation and allowing
more electrons to be active and to interact is important and leads to the generation of considerably more harmonics -- will remain unchanged. A full understanding of how two or more active electrons impact the HHG is a rather involved subject and goes much beyond the present work.

Nevertheless, we can say already now that having more discrete orbitals implies many more bound states of the two and in other systems possibly more electrons participating in the process. These bound states alone can also give rise to some HHG \cite{nimrod}.   In Ref.~\cite{nimrod} a realistic model for an array of quantum dots with six active correlated bound electrons has been solved numerically exactly and shown to unambiguously give rise to a second  plateau in the HHG spectrum. Although the situation in Ref.~\cite{nimrod} differs from ours, this result supports our finding that allowing more electrons to be active generates higher harmonics and is likely to give rise to a second plateau. In our present example of He, the ionization of an electron plays a crucial role, but the existence of the additional bound states certainly leads to additional different pathways of the important ionization channel and this will also influence the HHG. Finally, after one electron is ionized, the He$^+$  ion can stay in several excited states. The recombination step is then also different from that in the SAE approximation as there are many new pathways for it now.

\begin{acknowledgements}
This work was partly supported by the State Hessen initiative LOEWE within the focus-project ELCH, by the F\"{o}rderprogramm zur weiteren Profilbildung in der Universit\"{a}t Kassel (F\"{o}rderlinie \emph{Gro{\ss}e Br\"{u}cke}), by the Deutsche Forschungsgemeinschaft the within DFG project No. DE 2366/1-1, and by the U.S. ARL and the U.S. ARO under Grant No. W911NF-14-1-0383. Ph.V.D. acknowledges Research Institute of Physics, Southern Federal University for the hospitality and support during his research stay there.
\end{acknowledgements}

\end{document}